# SOLAR WIND DRIVEN PLASMA FLUXES FROM THE VENUS IONOSPHERE

H. Pérez-de-Tejada[1], R. Lundin[2], S. Barabash[2], T. L. Zhang[3], J. A., Sauvaud[4], H. Durand-Manterola[1], M. Reyes-Ruiz[5],

1 – Institute of Geophysics, UNAM, México, D. F.
2 – Swedish Institute of Space Physics, Kiruna, Sweden
3 – Space Research Institute, Graz, Austria
4 - CESR, Toulouse, France
5 – Institute of Astronomy, UNAM, Ensenada, México

Abstract. Measurements conducted with the ASPERA-4 instrument and the magnetometer of the Venus Express spacecraft show that the dynamic pressure of planetary O+ ion fluxes measured in the Venus wake can be significantly larger than the local magnetic pressure and, as a result, those ions are not being driven by magnetic forces but by the kinetic energy of the solar wind. Beams of planetary O+ ions with those properties have been detected in several orbits of the Venus Express through the wake as the spacecraft traverses by the noon-midnight plane along its near polar trajectory. The momentum flux of the O+ ions leads to superalfvenic flow conditions. It is suggested that such O+ ion beams are produced in the vicinity of the magnetic polar regions of the Venus ionosphere where the solar wind erodes the local plasma leading to plasma channels that extend downstream from those regions.

## Introduction

The acceleration of planetary ions from the Venus ionosphere has been extensively considered in relation to the observation of important features of the plasma environment of that planet. Measurements conducted with the Pioneer Venus Orbiter (PVO) revealed the existence of separate blobs of ionospheric plasma that have been interpreted as representing clouds or crossing through plasma filaments or tails that extend downstream from Venus

[Brace et al, 1982a]. At the same time the electron density profiles in the nightside ionosphere also show the presence of regions in which the plasma density decreases to very low values and that have been interpreted as ionospheric holes [Brace et al., 1982b]. While plasma clouds have been detected more frequently than the ionospheric holes there is a tendency for the latter features to occur in the inner wake within a region that is somewhat displaced towards the dawn side. The distribution of the ionospheric holes in the magnetic field oriented reference frame was examined by Marubashi et al. [1995] and has recently also been addressed by Hoegy and Grebowsky [2010].

The interpretation of plasma clouds has relied on the observation that the interplanetary magnetic field measured with the PVO wraps around them and that there is a change in the polarity of the sun-Venus component of the magnetic field vector as the spacecraft moves behind them and probes their cross section [Russell et al., 1982]. This geometry of the magnetic field has also been assumed to be applicable around and downstream from the magnetic polar regions of the Venus ionosphere where the field lines that have piled up around the dayside ionopause form a hairpin configuration to later enter the wake. The conditions in those regions are related as well to the acceleration of the local planetary particles that are driven into the wake. While magnetic forces should be involved in producing the motion of those particles it has not been possible to address their kinetic properties. Data obtained from the Venus Express spacecraft (VEX) have now provided information with which it is possible to examine this issue.

Different processes have been suggested for the interpretation of the ionospheric holes including electric currents directed along magnetic field lines that stretch along the tail [Grebowsky et al., 1981] or plasma erosion produced by the solar wind on the magnetic polar regions of the ionosphere [Perez-de-Tejada, 2004]. In this latter view the solar wind carves out a plasma polar channel or duct from those regions that stretch over the nightside hemisphere allowing for an open boundary to the ionospheric holes. A necessary condition required to support dynamic processes as the source of the plasma channels is that the kinetic energy density of the plasma by the polar regions (dynamic pressure) is larger than the local magnetic energy density (magnetic pressure) and that in this sense the solar wind is responsible for driving the plasma clouds. These questions will be addressed below in the analysis of the plasma and magnetic field data obtained from VEX.

## VEX data

A suitable example of measurements of the VEX plasma and magnetic field data is obtained during orbit 123 on August 22, 2006 that is reproduced in Figure 1. In this orbit the spacecraft moved towards the planet from the wake by the midnight plane (small Y values at the foot print of Figure 1) while it traveled from the south hemisphere to the north hemisphere (from negative to positive Z values). The upper two panels in Figure 1 describe the energy spectra of H+ and O+ ions measured by the Aspera-4 ion mass analyzer (IMA) as the spacecraft approached Venus and exited upstream from the planet (the nominal IMA energy scale from 2006 to July 2010 starts at 10 eV/q). As of July 2010 new energy settings were introduced, the IMA energy scale now starts at 1 eV/q. The new measurement results clearly demonstrate that the VEX spacecraft (s/c) is charged up to some -10V near pericenter and in the wake, thus motivating a -9V spacecraft potential used in the analysis of low energy ions [e.g. Lundin et al., 2011]. However, a detail analysis of the results presented here show that the difference between using nominal energy settings and a flat -9V s/c charging setting is less than 20 %. In order to avoid further dues on this matter we have used here nominal energy settings. Notable in the third panel of Figure 1 is the observation of significant densities of O+ ions as the spacecraft moved near the terminator (by 02:37 UT) and also during the inbound crossing of the wake from the south hemisphere (in the 01:37 UT – 01:58 UT time range). The velocity of the H+ and O+ ions shown in the fourth and fifth panels has been obtained after removing the spacecraft speed and there is a large contribution of the $V_x$ component in the measurements of the O+ ion fluxes along the flanks of the wake. As it has been inferred from the data of other orbits [Zhang et al., 2010] there is evidence in the bottom panel that near the terminator the spacecraft moved through a region where the magnetic field is piled up (enhanced values of the $B_y$ and $B_z$ components). At the same time it is significant that depressed (diamagnetic) values of the magnetic field were measured throughout the region where the O+ ions are detected in the inbound crossing of the wake (between 01:37 UT and 01:52 UT). In this sense it should also be noted that downstream from the region where the magnetic field is piled up (before 02:30 UT) its intensity maintains < 20 nT values and that it is further smaller at the time when the O+ ion fluxes are measured.

From the density and speed values of the H+ and O+ ions shown in Figure 1 it has been possible to derive the profiles of the dynamic pressure of both components along the trajectory of the VEX spacecraft in orbit 123 and that are presented in the lower panel of Figure 2. We have also included in that figure the profile of the magnetic field pressure to show that the dynamic pressure of the O+ ions becomes substantially larger than the magnetic pressure in the inbound crossing of the wake (between 01:48 UT and 02:00 UT). The dynamic pressure of the H+ ions is also larger than the magnetic field pressure upstream of the bow shock (which occurs by ~ 02:53 UT) and downstream from the region where the density and speed were of those ions were influenced by the plasma wake (before ~ 01:35 UT). However, within a wide region of the wake (bounded near ~ 02:00 UT and ~ 02:30 UT) the H+ and O+ ion density maintains low values and thus their local dynamic pressure becomes smaller than the magnetic pressure. It is also of interest to note that the peak value of the dynamic pressure of the O+ detected near the terminator (by ~ 02:37 UT) is comparable to that measured during the inbound crossing into the wake (by ~ 02:00 UT) suggesting that fluxes of that ion component preserve their kinetic energy density as they move downstream from the planet. At the same time while the magnetic pressure by the terminator is larger than the local dynamic pressure of the O+ ions it is found that in the near wake (by ~ 02:35 UT) the total dynamic pressure of the plasma fluxes (thick profile) becomes comparable to the local magnetic pressure thus suggesting that it can be sufficient to drive the O+ ions within the wake. The low values of the dynamic pressure of the H+ ions in the wake (between 02:00 UT and 02:30 UT) with respect to those present in the freestream solar wind derive from the manner in which their density decreases in that region as it is shown in the third panel of Figure 1, and that is suggestive of an overall expansion of the plasma.

An important consequence of the different values of the local magnetic pressure with respect to the dynamic pressure of the ion components in the Venus wake is to estimate the Alfven Mach number $M_A$ that they imply. The results of that calculation derived from the values presented in the lower panel of Figure 2 lead to the profile of the ratio of the total dynamic pressure to the magnetic pressure that is shown in the upper panel of that figure, and that exhibits conditions with important implications on the motion of the particles. In particular, different from the large variations of the pressure ratio it is significant that it leads to superalfvenic ($M_A > 1$) flow conditions (values above the horizontal line) in certain regions of the wake and, most notably, when the O+ ion fluxes were measured during the inbound pass of the spacecraft to the wake (in their peak value at ~ 01:50 UT we have $M_A \cong 4$). It is to be noted that the up to 20 % difference inferred from the IMA energy settings and that applies to the ion density values does not modify the superalfvenic conditions of the flow. At the same time since similar values of the Alfven Mach number were obtained for the H+ ion fluxes in the freestream solar wind (after the bow shock crossing) the main contribution of the results presented in the upper panel of Figure 2 is that the O+ ions in the wake have replaced the freestream solar wind H+ ions in the convection of the magnetic field. As it is the case in the freestream solar wind the motion of the O+ ions will now influence the distribution of the magnetic field fluxes which will remain attached to the plasma flow. Superalfvenic flow conditions are also present at peak values of the total dynamic pressure (at ~ 02:32 UT in the upper panel of Figure 2) or at low values of the magnetic field intensity but as a whole it is possible that the plasma and the magnetic field pressure participate jointly in the displacement of the particles.

Comparable variations of the dynamic and the magnetic field pressure along the VEX trajectory have been observed in the data of other orbits with values that resemble those shown in Figures 1-2. The plasma properties measured in orbit 132 on August 31, 2006 exhibits changes in the density and speed of the H+ and O+ ions that are similar to those observed in Figure 1. Profiles of the dynamic pressure of the H+ and the O+ ion components together with that of the magnetic pressure along the VEX trajectory for orbit 132 are shown in the lower panel of Figure 3. As it was the case for the data of orbit 123 in Figure 2 it can be appreciated that the dynamic pressure of the O+ ions by 02:19 UT is larger than the local magnetic field pressure thus substantiating the dominant value of the kinetic energy in their motion. While that peak value of the dynamic pressure of the O+ fluxes is nearly one order of magnitude larger than the local magnetic field pressure it is comparable to the magnetic pressure that is present where such ions are not observed (in the 02:35 UT – 03:00 UT time interval). This agreement may imply that the dynamic pressure of the O+ ions is sufficient to replace locally the magnetic pressure that is available in the wake, in other words that the O+ ions have been propelled from the ionosphere with kinetic energy density values that are equivalent to the magnetic energy density that is present in the region where they are not measured (the kinetic energy of the solar wind that reaches the polar terminator is employed to accelerate the O+ ions and at the same time is used to determine the value of the magnetic field intensity in the wake). The value of the pressure ratio $P_{p\_tot}/P_B$ in the upper panel of Figure 3 at the time when the intense O+ ion fluxes were measured during the inner crossing of the wake (in the 02:15 UT – 02:25 UT time interval) gives evidence that such ions move under superalfvenic flow conditions and that at other locations the plasma is mostly influenced by magnetic pressure.

The observation of plasma properties that lead to superalfvenic flow conditions for the O+ ion fluxes in the Venus wake provides a useful interpretation of the motion of the charged particles in that region. Rather than moving along the magnetic field lines the collective motion of the O+ ions is sufficient to modify the magnetic field geometry as the particles are driven by the kinetic energy of the solar wind. Magnetic tension forces should be applicable in the region where the plasma conditions are subalfvenic (as it is the case in the magnetic barrier) but dynamic forces will be more important than the magnetic forces when the superalfvenic O+ ion fluxes are measured in the wake. Dynamic forces will be responsible for producing the motion of those ions with conditions that can be better depicted from a fluid dynamic approach. Earlier calculations leading to superalfvenic flow conditions at the flanks of the Venus ionosheath where the solar wind has interacted with the Venus ionosphere were reported from measurements conducted with the Mariner 5 spacecraft [Pérez-de-Tejada, 1999].

Discussion

A schematic view of the manner in which the solar wind deforms the Venus ionosphere as it directly interacts with its magnetic polar regions is depicted in Figure 4. The solar wind can reach the upper ionospheric plasma in those regions because the local enhancement of interplanetary magnetic fluxes around that part of the ionosphere is less severe than what occurs elsewhere over the dayside ionopause (magnetic barrier). As a result the kinetic energy of the solar wind is used to carve out plasma channels by the magnetic polar regions and that extend over the nightside ionosphere. Crossings of those channels along the PVO trajectory were interpreted as leading to the observation of ionospheric holes with conditions in which the observation of those features depends on the region that is probed by that spacecraft [Pérez-de-Tejada, 2004]. The low (~ 140 km) altitude of periapsis of the PVO trajectory allowed crossings through the north hemisphere polar channel, and leads to different observations conducted with the VEX spacecraft in which the periapsis of its trajectory occurs at higher (~ 300 km) altitudes. A prominent result is the identification of low energy planetary O+ ion fluxes that have been eroded from the ionosphere by the solar wind [Barabash et al., 2007]. The peak values of the O+ ion fluxes measured during the inbound pass into the wake in the southern hemisphere, by 01:50 UT in Figure 1, occurred before the spacecraft approached the planet over the north hemisphere. Those O+ ions are mostly detected when the VEX spacecraft scans downstream from the magnetic polar region that, in turn, is identified by a change in the polarity of the $B_x$ component by 02:37 UT in Figure 1.

The geometry of the plasma channels is determined by the position of the magnetic polar regions which in turn is dictated by the orientation of the IMF with respect to the ecliptic plane in the freestream solar wind. From the magnetic field data of Figure 1 it can be expected that the magnetic polar regions are not placed by the geographic poles but along the line that is traced between them and that is displaced ~45° from the z-axis at the terminator plane (the $B_y$ and the $B_z$ components upstream from the bow shock exhibit nearly equal values and thus lead to that angle). Under such circumstances the plasma channels in Figure 4 are tilted away from the z-axis and thus are located in a region that may not have been probed along the polar oriented trajectory of the Venus Express. This may have been the case for measurements made in the north hemisphere through the wake (positive z values) where as shown in Figure 1 very weak O+ ions fluxes were only detected by ~ 02:15 UT before the spacecraft reached the terminator.

The erosion and removal of planetary O+ ions by the kinetic energy of the solar wind are processes that lie at the base of an interpretation that is required to understand their physical properties. Rather than assuming magnetic forces to account for the motion of those ions it is necessary to implement dynamic forces as mostly responsible for the manner in which they are accelerated. Wave particle interactions implied by the strong fluctuations of the magnetic field that have been measured in the Venus wake [Bridge et al., 1967; Voros et al., 2010] should be ultimately responsible for the dynamic processes that lead to the transport of solar wind momentum to the Venus upper ionosphere. Calculations of the momentum flux of planetary ions measured near periapsis along the VEX trajectory show that it is comparable to the momentum flux of the freestream solar wind protons thus supporting that an efficient transport of momentum takes place between both ion populations [Pérez-de-Tejada et al., 2011]. The present results now show the small contribution of the magnetic field forces to locally accelerate the solar wind driven planetary ions whose kinetic energy will end up being delivered to the neutral atmosphere through particle-particle interactions and thus help support its retrograde rotation [Lundin et al., 2011].

We have examined the corresponding profiles for other VEX orbits in 2006 and during 2009 in which the spacecraft also probed the noon-midnight plane and found similar differences in the value of the dynamic and the magnetic pressure presented in Figures 2 and 3. As it was the case for the two orbits selected in the present study we find that the O+ ion fluxes measured in other orbits also acquire energies that lead to superalfvenic flow conditions in the Venus wake thus implying that the properties indicated here appear to be persistent. A more extended discussion of the distribution of the region where the O+ ions beams are measured, together with estimates of the width of those regions and their correlation with the solar wind will be presented in a separate study.

Acknowledgments. The research conducted in this study was supported by Project IN110407 of DGAPA at UNAM in México. The authors wish to thank Gilberto. Casillas at UNAM in México D. F., and Pär-Ola Nilsson and Fredrik Rutqvist at IRF in Umeå, Sweden, who provided technical assistance. Finally, a special thanks to the National Space Agencies in Sweden, France, and Austria for supporting the ASPERA and MAG instruments, and to the European Space Agency for making VEX a great success.

Figure captions

Fig.1 – Energy spectra of $H^+$ and $O^+$ plasma fluxes (first and second panels) measured in the Venus wake and over the north polar region during orbit 123 of the Venus Express spacecraft in August 22, 2006. Density profiles of both plasma components (third panel) and their speed V, together with their $V_x$, $V_y$, and $V_z$ velocity components are shown in the fourth and fifth panels. The bottom panel shows the magnetic field profiles (magnitude and components) measured along the same orbit.

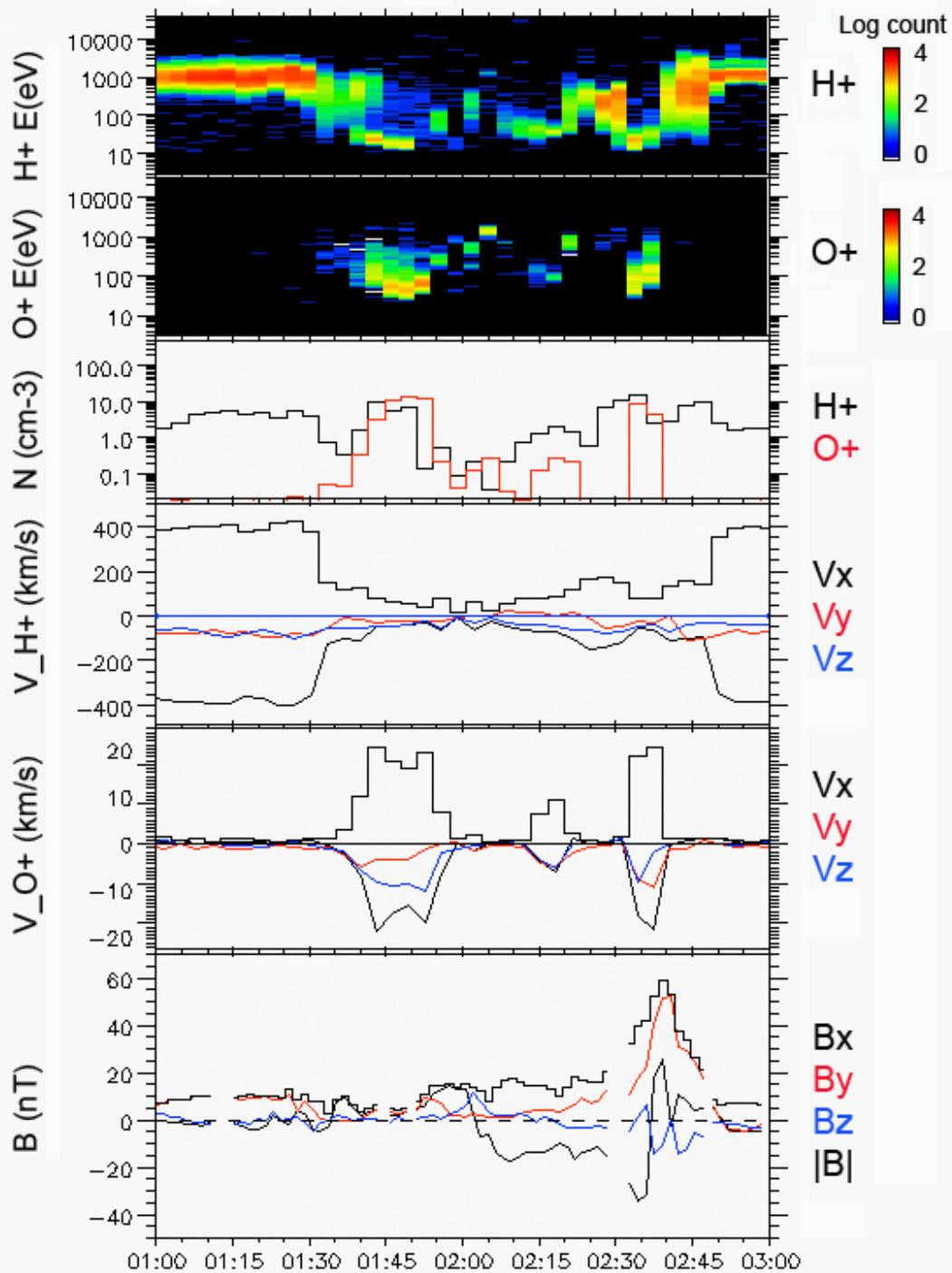

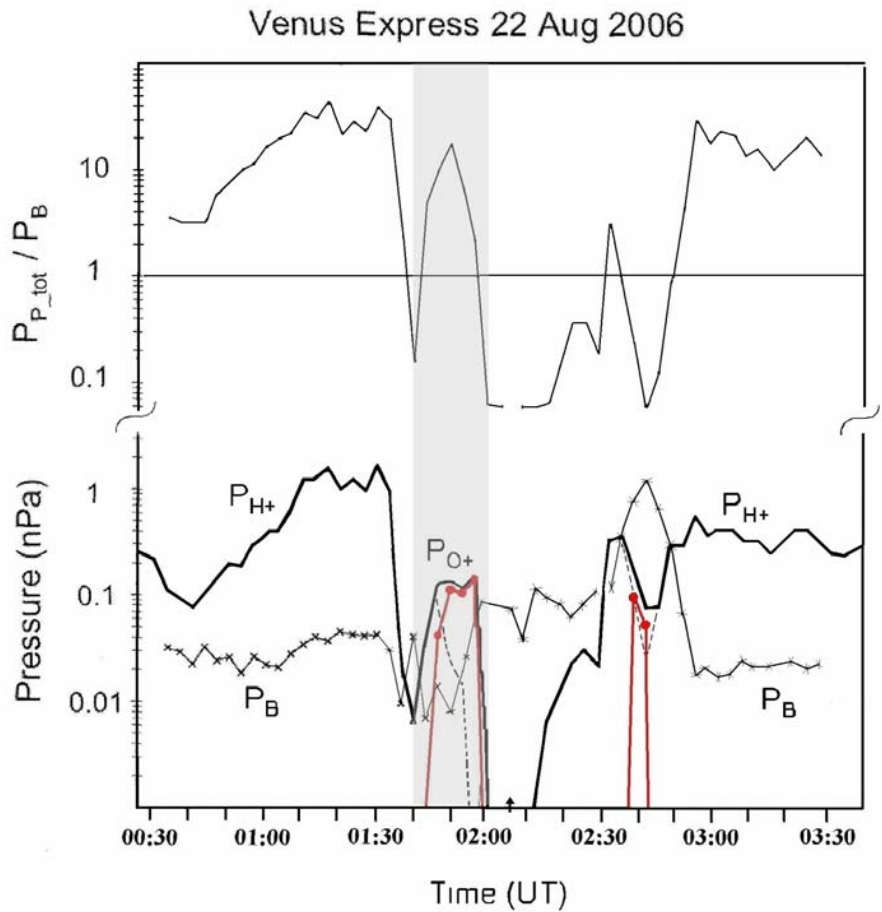

Fig. 2 – (lower panel) Profiles of the dynamic pressure of the O+ ions (marked in red) and the H+ ions, together with the profile of the magnetic field pressure measured through the Venus wake during orbit 123 of the Venus Express spacecraft in August 22, 2006 (the dotted profile at the time when the O+ ions are detected is the dynamic pressure of the H+ ions, and the heavy profile above it gives the added value of the dynamic pressure of both components). (upper panel) Ratio values of the total dynamic pressure of the plasma to the magnetic field pressure derived from the profiles shown in the lower panel. The outbound bow shock crossing occurs at ~ 02:53 UT and the peak value at ~01:50 UT is provided by the dynamic pressure of the O+ ions).

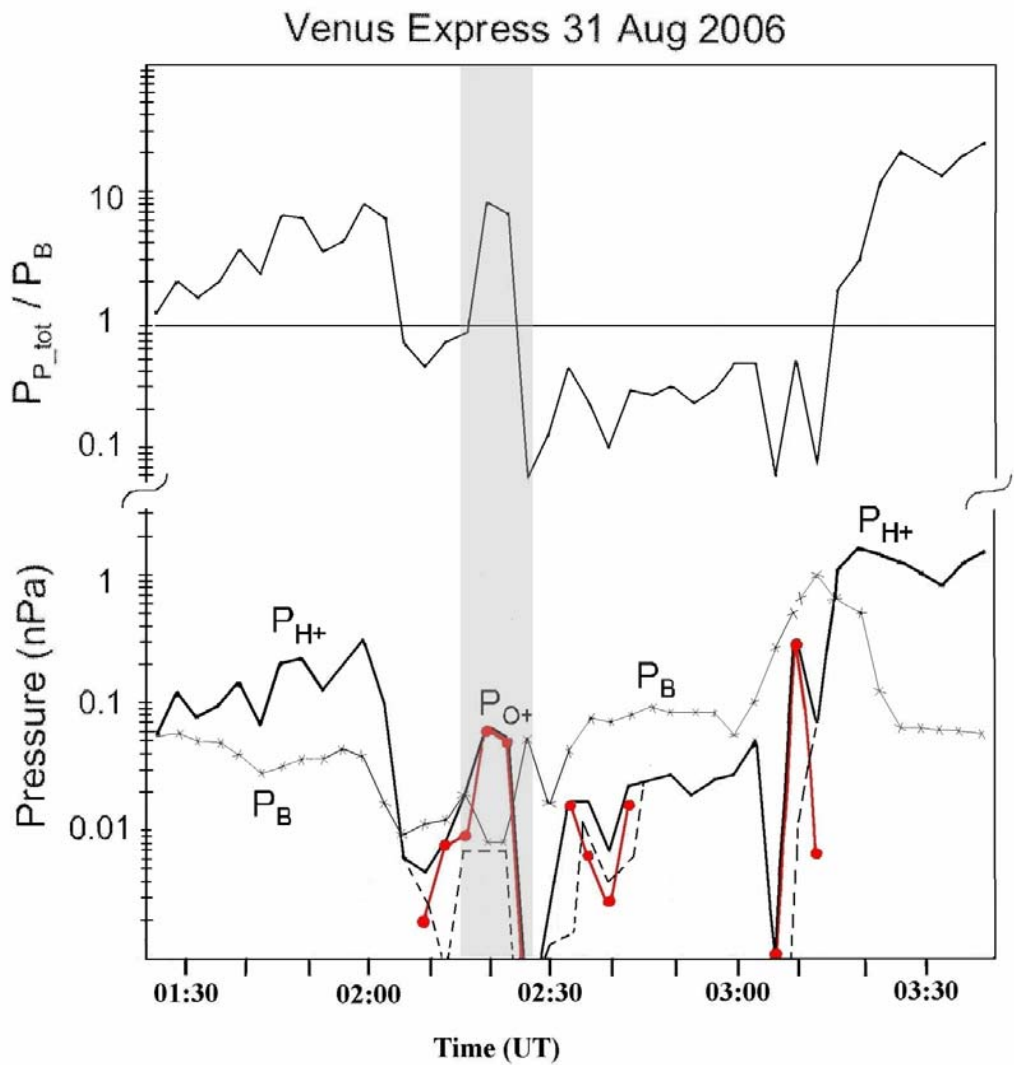

Fig. 3 – (lower panel) Profiles of the dynamic pressure of the O+ ions and he H+ ions, together with the profile of the magnetic field pressure that were measured through the Venus wake during orbit 132 of the Venus Express spacecraft in August 31, 2006. (upper panel) Ratio values of the total dynamic pressure of the plasma to the magnetic field pressure derived from the profiles shown in the lower panel. The outbound bow shock crossing occurs at 03:22 UT and the peak value at 02:19 UT is provided by the dynamic pressure of the O+ ions).

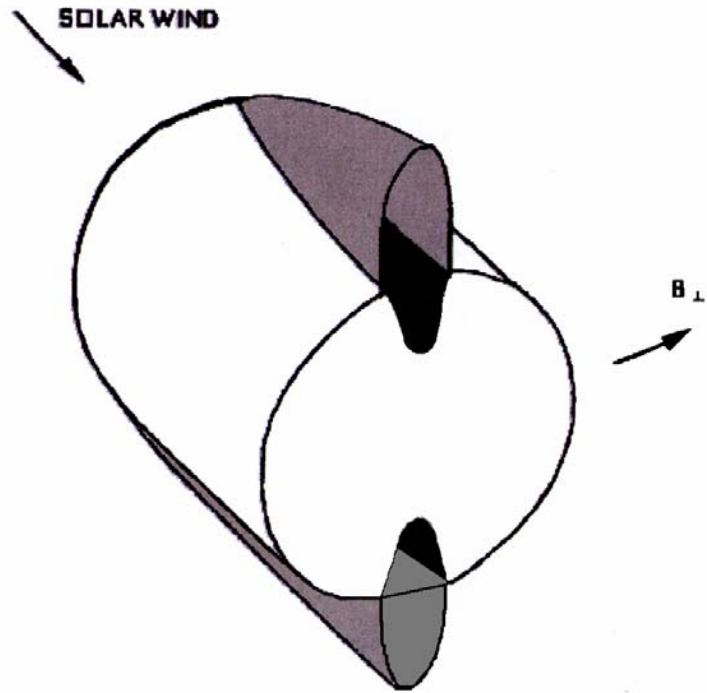

Fig. 4 – Schematic view of plasma channels that form by and downstream from the magnetic polar regions of the Venus Ionosphere together with the region above them where planetary ions are eroded by the solar wind [after Pérez-de-Tejada et al., 2009].